\documentclass[sigconf]{acmart}

\usepackage{color}
\usepackage{float}
\usepackage{graphicx}
\usepackage{subfig}
\usepackage{multirow}
\usepackage{bbding}
\usepackage{booktabs}
\usepackage{amsmath}
\usepackage{breqn}
\usepackage{mathrsfs}
\usepackage[ruled]{algorithm2e}
\usepackage{hyperref}
\usepackage[normalem]{ulem}

\usepackage{colortbl}

\graphicspath{{img/}}

\AtBeginDocument{%
  }

\setcopyright{acmlicensed}
\copyrightyear{2023}
\acmYear{2023}
\acmConference[KDD '23] {Proceedings of the 29th ACM SIGKDD Conference on Knowledge Discovery and Data Mining}{August 6--10, 2023}{Long Beach, CA, USA.}
\acmBooktitle{Proceedings of the 29th ACM SIGKDD Conference on Knowledge Discovery and Data Mining (KDD '23), August 6--10, 2023, Long Beach, CA, USA}
\acmPrice{15.00}
\acmISBN{979-8-4007-0103-0/23/08}
\acmDOI{https://doi.org/10.1145/3580305.3599252}

\begin{document}

\title{Automated 3D Pre-Training for Molecular Property Prediction}
%\title{Adaptive 3D Pre-Training for Molecular Property Prediction}

\author{Xu Wang}
\email{wangxu01@4paradigm.com}
\affiliation{%
	\institution{4Paradigm Inc.}
	\city{Beijing}
	\country{China}
}

\author{Huan Zhao}
\authornote{Huan is the corresponding author.}
\email{zhaohuan@4paradigm.com}
\affiliation{%
	\institution{4Paradigm Inc.}
	\city{Beijing}
	\country{China}
}

\author{Wei-wei Tu}
\email{tuweiwei@4paradigm.com}
\affiliation{%
  \institution{4Paradigm Inc.}
  \city{Beijing}
  \country{China}
}

\author{Quanming Yao}
%\authornotemark[2]
\email{qyaoaa@tsinghua.edu.cn}
\affiliation{%
	\institution{Department of Electronic Engineering, \\ Tsinghua University}
	\city{Beijing}
	\country{China}
}


\renewcommand{\shortauthors}{Xu Wang, Huan Zhao, Wei-wei Tu, \& Quanming Yao}

\begin{abstract}

Molecular property prediction is an important problem in drug discovery and materials science. 
As geometric structures have been demonstrated necessary for molecular property prediction,
3D information has been combined with various graph learning methods
to boost prediction performance.
%
%Various methods, including graph neural networks and graph transformers, 
%have been proposed  
%
%for 
%this task as a molecule can be naturally represented by a graph.
%Besides, geometric structures have been demonstrated necessary for better performance of molecular property prediction. 
However, obtaining the geometric structure of molecules 
is not feasible in many real-world applications due to the high computational cost.
%\footnote{+qm+ what is the ``challenging problem''?}
In this work, we propose a novel 3D pre-training framework (dubbed 3D PGT), 
which pre-trains a model on 3D molecular graphs, and then fine-tunes it on molecular graphs without 3D structures.
Based on fact that \textit{bond length}, 
\textit{bond angle}, and \textit{dihedral angle} are three basic geometric descriptors corresponding to a complete molecular 3D conformer, 
we first develop a multi-task generative pre-train framework based on these three attributes.
Next, to automatically fuse these three generative tasks, 
we design a surrogate metric using the \textit{total energy} to search for weight distribution of the three pretext tasks
since total energy corresponding to the quality of 3D conformer.
%The proposed framework is built on top of a graph transformer backbone, thus it is dubbed 3D PGT (Pretrained Graph Transformer).
Extensive experiments on 2D molecular graphs are conducted to demonstrate the accuracy, 
efficiency and generalization ability of the proposed 3D PGT compared to various pre-training baselines. 
%Especially, 3D PGT outperforms all baselines from top solutions for large-scale molecular prediction in OGB leaderboard.

\end{abstract}






\maketitle
\vspace{-8pt}
\section{Introduction}

\begin{figure}[t]
	\centering
	\includegraphics[width=\linewidth]{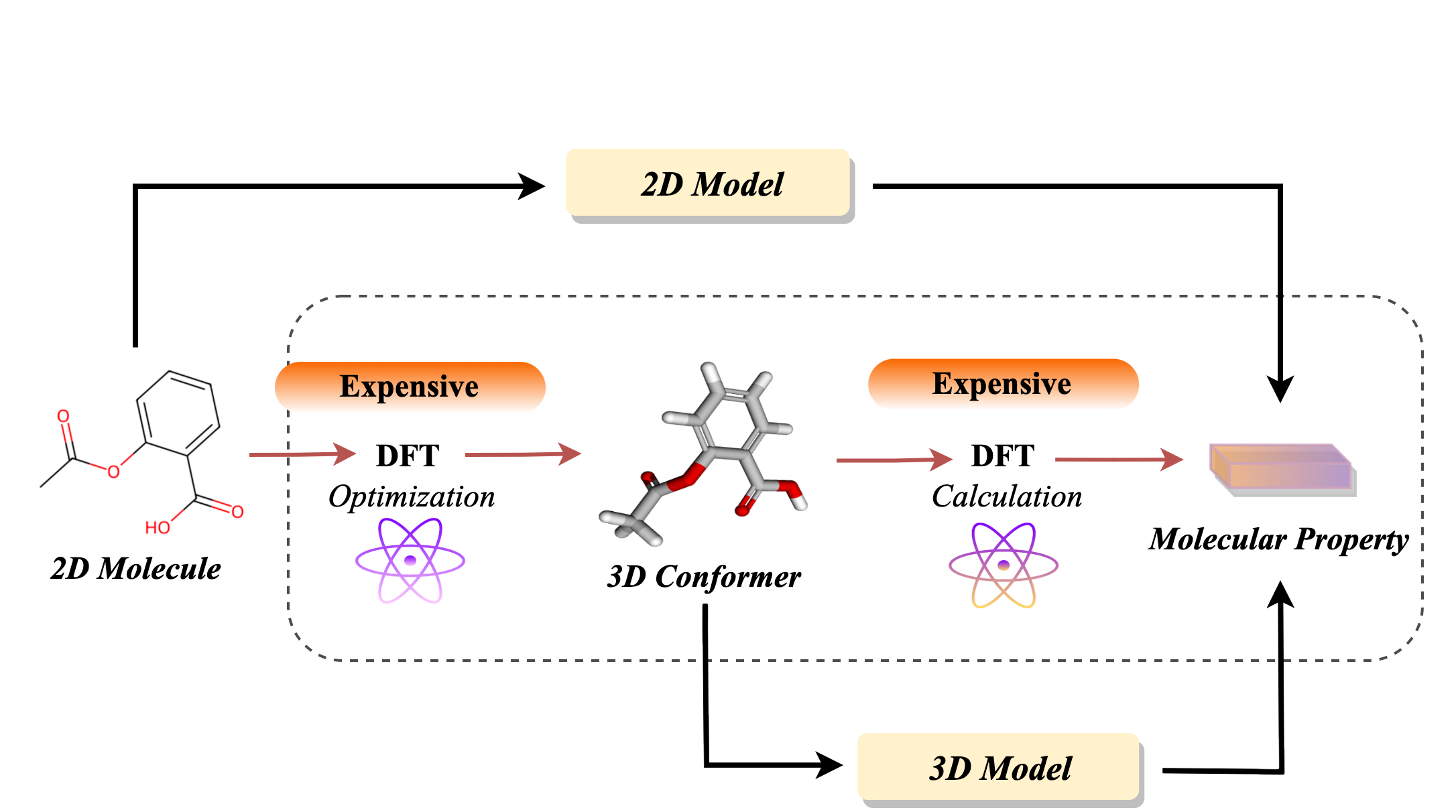}
	\caption{The general pipeline of existing methods for molecular property prediction. The DFT-based method first optimizes the molecular 3D structure to obtain a low-energy conformer, and then calculates specific molecular properties based on the conformer. The 2D model approximates the whole process of DFT, and directly predicts the properties through the input 2D molecules, while the 3D model requires specific geometric information.}
	\label{fig:background}
\end{figure}

Molecular property prediction is a foundational problem in drug discovery \cite{stokes2020deep}, materials science \cite{chanussot2021open, tran2022open} and bioinformatics \cite{narayanan2002artificial, wang2022graph, zhou2023unimol}.
Conventionally, in quantum chemistry, popular computational tools based on Density Functional Theory (DFT) \cite{parr1983density} are used to compute a molecule's geometry and quantum properties, e.g. energy of atoms.
Then recent years have witnessed the success of machine learning, especially graph neural networks (GNNs)  \cite{gilmer2017neural,faber2017machine} and graph transformer based methods \cite{ying2021transformers, wu2021representing, hussain2022global}, in molecular property prediction,
since a molecule can be naturally represented by a graph, where nodes and edges represent atoms and their chemical bonds, respectively.
As shown in Figure~\ref{fig:background}, existing learning-based methods can be roughly categorized into two groups based on the usage of 3D geometry of molecules. 
Especially, as the molecular properties are mostly determined by their 3D structures \cite{crum1865connection, hansch1964p}, methods \cite{gasteiger2021gemnet, klicpera2020directional, zitnick2022spherical} that directly model the 3D structures of molecules, e.g., 3D conformers, have been proposed to further improve the performance compared to 2D methods.
However, due to the high computational expensive of DFT-based tools, it tends to be a challenging problem to obtain sufficient labels for these learning based methods.

Very recently, inspired by the success of pre-training and finetuning paradigm in natural language processing \cite{devlin2018bert,radford2018improving} and computer vision \cite{chen2020simple,he2022masked},
researchers started to apply self-supervised learning (SSL) to molecular property prediction \cite{hu2019strategies,hu2020gpt,qiu2020gcc,zaidi2022pre,liu2022pre,stark20223d}. The
key to pre-training lies in pretext tasks design to make better use of the rich unlabeled molecule data, no matter 2D or 3D graphs. Then, existing works can also be categorized into 2D and 3D pre-train methods,
according to the type of pretext tasks. For example, in \cite{hu2019strategies}, several node-level and graph-level prediction tasks directly from 2D molecular graphs are designed to pre-train GNN models, which are then fine-tuned on downstream tasks. In terms of 3D pre-training methods, \cite{zaidi2022pre} conducts SSL tasks by position denoising of 3D molecular structures at equilibrium to learn the molecular force field, which can help the 3D model predict properties more accurately.  All these works have demonstrated the effectiveness of pre-training methods for molecular prediction, especially designing pretext tasks
on 3D molecular structures.

Although 3D molecular structures are necessary to achieve strong performance in molecular property prediction \cite{frenkel2001understanding}, 
obtaining 3D equilibrium structure, shown in Figure \ref{fig:background}, still requires expensive DFT-based geometry optimization \cite{hu2021ogb, xu2021learning}. Thus, even if we can obtain an accurate 3D GNN model by pre-training, prediction on downstream tasks can still be very slow, especially for large-scale real-world applications \cite{hu2021ogb}, since we need to compute 3D structure for each molecule before inference.

In this work, to address this problem, we propose a novel 3D pre-train approach on existing unlabeled 3D molecular datasets and then fine-tune and predict on the molecules \textit{only with 2D molecular graphs.} In this way, the proposed method can enjoy the benefits of both two worlds, i.e., high accuracy from pre-training on 3D molecular graphs, and high efficiency from removing the need for computing 3D structures in the fine-tuning/prediction stage.
To be specific, we first design multiple generative pre-training tasks based on three attributes, \textit{bond length}, \textit{bond angle}, and \textit{dihedral angle}, which are three basic local geometric descriptors that can be combined into a complete molecular 3D conformer\cite{mcnaught1997compendium}.
In this work, we build our 3D multiple pretext tasks on the assumption that these three tasks are necessary and sufficient for effective encoding of 3D structure in molecular property prediction, then it remains to be a critical problem that fuse these three tasks in the pre-training stage.
Intuitively, different tasks have distinct importance given a pre-trained dataset, and improper integration of multiple tasks may lead to ineffective or negative transfer to downstream tasks \cite{sun2022does}.
A possible solution is to search for an automated fusion scheme for these three pretext tasks, while the difficulty is that no supervision signal from downstream tasks can be provided.
To tackle this problem,  we design a surrogate metric based on \textit{total energy} of all atoms in a molecule in pre-training stage that can supervise the weight allocation of each pre-text tasks, considering that it corresponds to a lowest-energy 3D conformer for a molecule~\cite{mcnaught1997compendium, axelrod2022geom}.
By automatically searching for the pre-task and optimizing the total energy in the pre-training stage, we can potentially encode more accurate molecular geometry information on the 2D molecular graph, which will promise generalization benefits for various molecular property prediction tasks.
Besides, 
to guarantee the prediction performance, 
we choose GPS~\cite{rampavsek2022recipe} as our backbone,
which is a hybrid architecture of GNN and Transformer. 
Then the proposed method is dubbed 3D PGT (Pre-trained Graph Transformer).

To demonstrate the superiority of the proposed 3D PGT, 
we conduct extensive experiments on various scenarios. Here we highlight some results: 
a) On the well-known benchmark dataset, QM9 \cite{ramakrishnan2014quantum}, 
3D PGT not only improves the prediction performance compared to all baselines, 
but also enjoys high efficiency the same as methods directly operating on 2D molecular graphs; 
b) The performance gains in three kinds of downstream tasks demonstrate the strong generalization ability of the proposed 3D PGT; 
c) In a large-scale application on HOMO-LUMO energy gap prediction in OGB benchmark \cite{hu2021ogb} (3.37 million molecules for pre-training),
 where only 2D methods are feasible, the proposed 3D PGT achieves 10.6\% relative MAE reduction compared to the same backbone without pre-training and outperforms all models from top solutions in the OGB leaderboard.

To summarize, the contributions of this work are as follows:

\begin{itemize}
%     \item To the best of our knowledge, We propose 3D PGT, a novel adaptive fusion framework for 3D molecular pre-training which can obtain an effective pre-trained 2D graph encoder with implicit accurate 3D geometric prior and benefit on 2D molecular downstream tasks. 
      \item We propose a novel pre-training framework to encode 3D geometry with three important pretext tasks for molecular property prediction.
      To the best of our knowledge, we are the first to integrate these three tasks in a unifying manner for pre-training methods.

%    \item To the best of our knowledge, 3D PGT is the first 3D generative pre-training framework on the hybrid backbone of MPNN and Transformer for improving performance on 2D molecular graphs.\huan{The first (key) contribution tend to be conceptual or problem novelty, thus it should be ``the explore of this important problem by a pretrain framework''}
    \item Further, based on important quantum chemistry knowledge, we design a surrogate metric by \textit{total energy}, which serves as a supervision signal to search the adaptive weight of each pre-training task, and generalize the benefits to various downstream tasks. 
     
%     \item Experiments showing the performance advantages of 3D PGT compared with other pre-training methods in 8 downstream regression tasks and 7 classification tasks.
     
     \item Extensive experiments on various tasks demonstrate the superiority of the proposed 3D PGT in terms of accuracy, efficiency and generalization ability. 
\end{itemize}

\section{Related Work}
%\huan{Good to cite \cite{hu2020gpt,qiu2020gcc} as background info in Intro or Rel works for KDD submission.}

%\huan{
%For related work, the current organization is OK. I have some suggestions:
%
%$\bullet$ For Sec 2.1, we can introduce DFT, 2D (GNN, GT) and 3D methods (3D GNN). Remember to emphasize the drawback of expensive computation of 3D methods, which means the importance of our work.
%
%$\bullet$ For pretraining methods, besides the introduced works, try to point out the drawback, e.g., potential negative transfer, no good supervision signal of existing works (**Double check to make sure this claim is right**)
%}

%{
%\color{blue}
%+qm+ is it better to organize as
%\begin{itemize}
%\item 
%\end{itemize}
%}

%\footnote{+qm+ check comments at the intro:
%	rename this subSection~as ``Learning Representations for Molecular Graph''?}

\subsection{Property Prediction on Molecular Graph}

For property prediction on molecular graph, GNN has become an emerging research field which process it as a graph-level prediction task \cite{wang2022graph, sun2023graph}. 
%\footnote{+qm+ the ref format means you need to take it
%	as a subject not object.
%	\label{ft:2}}
After \cite{faber2017machine} demonstrated that ML models could be more accurate than hybrid DFT if explicitly electron correlated quantum data was available, 
%\footnote{+qm+ same here, check footnote~\ref{ft:2}.}
\cite{gilmer2017neural} proposed a message passing framework for chemical prediction problems that are capable of learning their own features from molecular graphs directly, and was considered as the most general architectures for realizing GNNs. However, message passing based methods is weak in capturing long-range dependencies \cite{rampavsek2022recipe}. Complementary, the Graph Transformer based methods represented by Graphormer \cite{ying2021transformers} which is directly build upon the standard Transformer realized the interaction between long-range nodes. Furthermore, EGT \cite{hussain2022global} uses global self-attention to update both node and edge representations, allowing unconstrained dynamic long-range interaction between nodes and improves the performance for the quantum-chemical regression task on the PCQM4Mv2 \cite{hu2021ogb} dataset. GraphTrans \cite{wu2021representing} and GPS \cite{rampavsek2022recipe} regard GNN and Transformer as complementary, and explore the coexistence framework between them. 

%\footnote{+qm+ note that I split this part out.}
At the same time, one important branch of improvement is to effectively use the geometric information contained in 3D conformers \cite{townshend2019end, axelrod2022geom}. 
There is a series of work focus on using relative 3D information which can be derived based on absolute 
Cartesian coordinates, 
such as bond length and bond angle \cite{klicpera2020directional}. 
GemNet \cite{gasteiger2021gemnet} further capture the information from dihedral angle to define all relative atom positions uniquely. Not limited in Cartesian coordinates, SphereNet \cite{liu2021spherical} propose a generic framework of 3D graph network (3DGN), and design the spherical message passing to realize 3DGN in the spherical coordinate system. 
%\footnote{+qm+ check footnote~\ref{ft:1}.}
However, although 3D information has been proved to be valuable for the prediction of molecular properties, the generation of 3D conformer requires DFT-based geometric optimization which needs expensive calculation. This makes it infeasible to explicitly calculate the molecular structure for large-scale datasets.

\begin{figure*}[t]
	\centering
	\includegraphics[width=\linewidth]{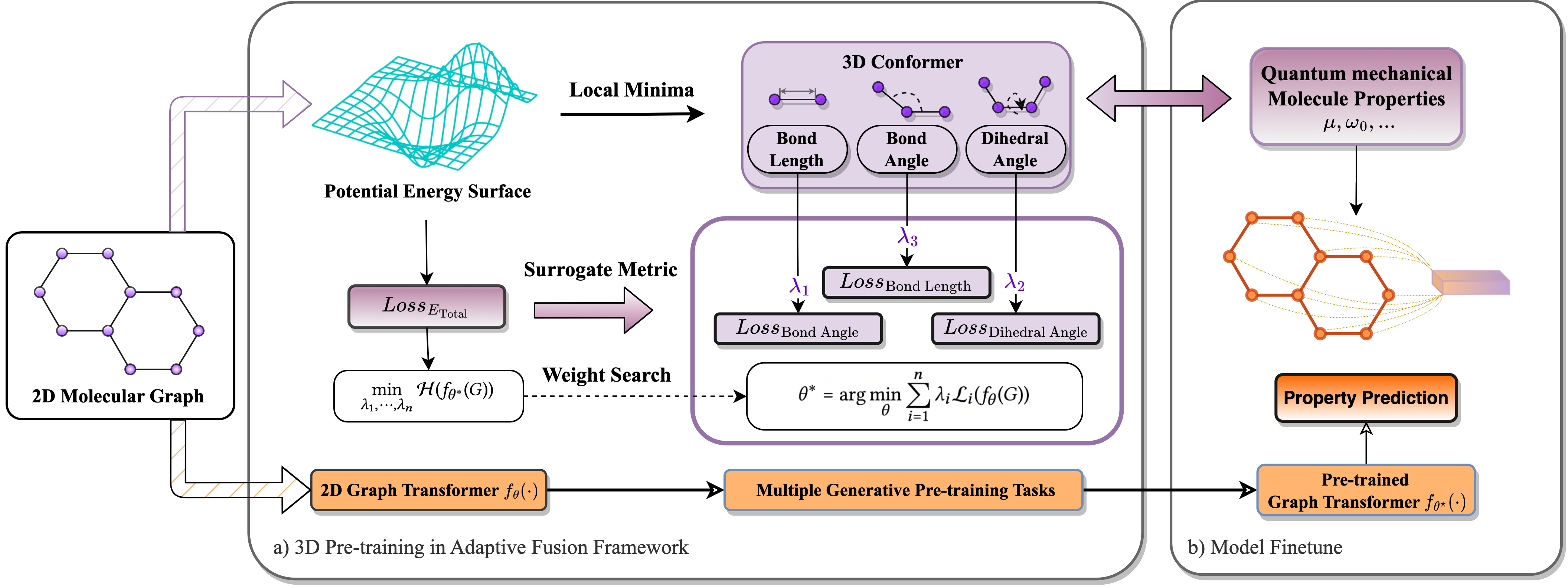}
	\caption{The overview of our Pre-training Framework. a) In the pre-training stage, we first design three generative pre-training tasks based on the low-energy conformer of the pre-training molecules, which are the prediction of \textit{bond length}, \textit{bond angle} and \textit{dihedral angle}. At the same time, we design a surrogate metric using the total energy corresponding to the low-energy conformer, and use it to search for the fusion weights of the three pre-training tasks. b) In the finetuning stage, since molecular geometry is valuable for its property prediction, the generation task we designed can introduce geometric priors to $f_{\theta}$, and benefit in downstream property prediction tasks.}
	\label{fig:framework}
	\vspace{-10pt}
\end{figure*}

\subsection{Pre-training on Molecular Graph}
%\huan{Then we may remove the ``3D'' in this title, and give review on typical pre-training works, e.g., 2D and 3D pre-training ones.}

%\footnote{+qm+ indeed this is the general pipeline of
%	pre-training,
%	however,
%	here,
%	the information used in the training and testing is not the same.
%	you need to explicitly emphasize this.}
The general pipeline of pre-training is to first pre-train the model on a large dataset through proxy tasks, and then conduct further parameter fine-tuning on specific downstream tasks based on corresponding supervision signals. 
The self-supervised pre-training on graph is usually designed to learn self-generated targets from the structure of the molecules, such as masked node reconstruction and context prediction \cite{hu2019strategies}. 
There are also works use motif prediction as a pre-training task on 2D molecular graph to improve the performance of property prediction \cite{rong2020self}.

Considering the calculation cost of 3D geometry and its importance for molecular property prediction task, a series of work proposed to pre-train a graph encoder which encode implicit 3D information in its latent vectors by pre-training on publicly available 3D molecular structures. They first conduct pre-training on the dataset with 3D structure through designing proxy task, and then transfer the pre-trained model to the downstream task which only contained 2D topology for fine-tuning. Similar to the 3D models which contained 3D geometry as input, the designed pre-training task designed should also respect the symmetries of rotation and translation. 
For example, Liu et al. \cite{liu2022pre} and St\"{a}rk et al. \cite{liu2022pre} used mutual information between 2D and 3D views for molecular pre-training, therefore the GNN is still able to produce implicit 3D information that can used to inform property predictions. 
Zhu et al. \cite{zhu2022unified} proposed a unified 2D and 3D pre-training which jointly leverages the 2D graph structure and the 3D geometry of the molecule. 
GEM \cite{gasteiger2021gemnet} proposes a bond-angle graph and self-supervised tasks which use large-scale unlabelled molecules with coarse 3D spatial structures which can be calculated by cheminformatics tools such as RDKit.
Inspired by the encoding method of Graphormer for 2D molecular graph, Transformer-M \cite{luo2022one} further encodes 3D spatial distance into attention bias, which can take molecular data of 2D or 3D formats as input.

Different from existing works, we design multiple generative pretext tasks based on 3D conformer, 
and fuse these tasks automatically by an interesting and useful supervision signal, i.e., the total energy of molecules.

\section{Method}

%\subsection{Problem Formulation}

\subsection{Problem Formulation} 
% 介绍清楚 Pre_training的 Pipeline，输入和输出分别是什么
\label{section:3.1}

%\subsection{Problem Setup}

Let $G_{\text{2D}}={(V,E)}$ denote a \textbf{2D molecular graph} with atom attributes $\mathbf{x}_{v}$ for $v \in V$ and bond attributes $\mathbf{e}_{uv}$ for $(u,v)\in E$, \textbf{3D molecular graph} refers to $G_{\text{3D}} ={(V,E,\mathbf{P})}$ which contains 3D Cartesian coordinates $\mathbf{P} \in \mathbb{R}^{\left|V\right| \times 3}$ for all atoms. The coordinates $\mathbf{P}$ contained in the 3D conformer is valuable information for the graph encoder $f_{\theta}(\cdot)$ in molecular property prediction, each conformer requires the calculation by DFT-based tools for several hours per CPU-core \cite{stewart2007stewart}. 

As generic pre-training pipelines, 3D pre-training with geometry for molecular representation has two stages:  
In the pre-training stage, the proxy task $\mathcal{L}_{\text{pre}}$ is designed to learn the self-generated targets from the publicly available 3D molecular graph $G_{\text{3D}}$; 
%\footnote{+qm+ good to illustrate this in the figure.}
then during the finetune stage, the pre-trained encoder $f_{\theta^{*}}(\cdot)$ is finetuned according to specific supervision signals on downstream tasks, which only includes 2D molecular graphs. 
 
%\subsection{}

%\huan{
%For the organization of this section, you can refer to the style of 3D infomax and GraphMVP, a possible version can be as follows:
%
%$\bullet$ An outline paragraph to introduce the content of Section~3. (see the example in the beginning paragraphs in Section~4 of 3D infomax paper.)
%
%$\bullet$ Some preliminary knowledge. No need to introduce GNN/Transformer. Let us discuss what to put here later.
%
%Good to give a formal problem setup. see the example in Sec 3.1 in \cite{zaidi2022pre}.(Btw, this should be a related work.)
%
%$\bullet$ An overview framework to introduce the whole process.
%
%$\bullet$ Adaptive fusion of multiple tasks. (you can also introduce the improved variant by RGC.)
%
%$\bullet$ (Optional) Some discussions, including multiple conformers, whether or not search for combination of GNN and Transformer, etc.
%}

\subsection{Automated Fusion Framework for Multiple 3D Pre-training Tasks} 
\label{tab:ourmethod}
%\huan{I do not think it necessary to introduce GNN and Transformer in 2023, since they have been quite popular in the past 5 years. In terms of preliminary, we may introduce sth about }
%\subsubsection{Fusion framework for multi-task pre-training}\

In this work, we propose 3D PGT, an automated fusion framework that contains multiple generative 3D pre-training tasks to obtain an effective pre-trained 2D graph encoder $f_{\theta^{*}}(\cdot)$. 
3D PGT focuses on 1. Designing multiple generative pre-training tasks that can benefit downstream tasks where 3D information is not available; 2. Bridging the gap between 3D generative pre-training tasks and downstream property prediction, allowing a wider range of downstream tasks to benefit from pre-training. 

In the following, we first present an overview of 3D PGT, and propose three generative pre-training tasks concerning 3D conformer generation. Then, to better fuse multiple pre-training tasks instead of purely average pre-training, we introduce an automated fusion framework with a surrogate metric to search the weight of each generative task during the pre-training stage. Finally, we summarize the variants for leveraging multiple conformers and searching the architecture of the backbone.

\begin{figure}[t]
 	\centering
 	\includegraphics[width=0.9\linewidth]{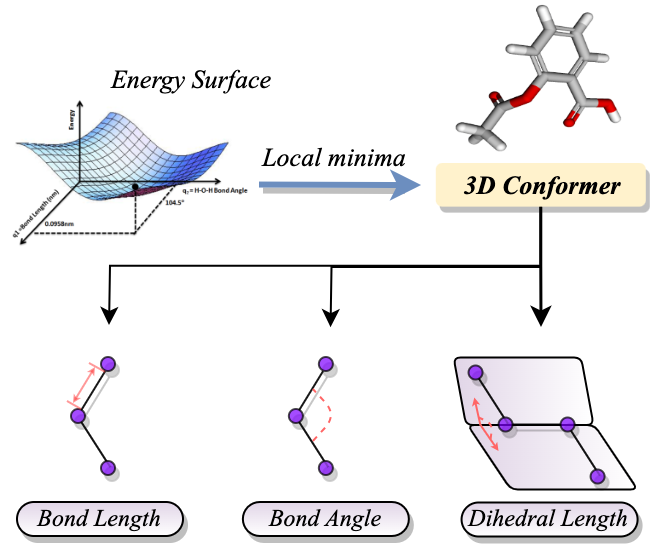}
 	\caption{A potential energy surface (PES) describes the energy of a system, defining the energy as a function of one or more coordinates. The optimization of the 3D conformer is to find the local minimum on the PES. At the same time, the geometric information contained in the 3D conformer can be described by the bond lengths of two connecting atoms, the bond angles of three connecting atoms, and the dihedral angles of three consecutive bonds.}
 	\label{fig:generative}
 \end{figure}

\noindent\textbf{Overview of 3D PGT.}
In general, 3D PGT focuses on designing the 3D pre-training framework to learn the robust transferable knowledge from $G_{\text{3D}}$ and then generalize to downstream tasks which consist of $G_{\text{2D}}$. 
Different from existing contrastive learning based methods \cite{liu2022pre, stark20223d}, 3D PGT aims to encode geometric priors only through single-model pre-training. 
As shown in Figure~\ref{fig:generative}, 3D PGT first splits the 3D conformer optimization into three generative tasks: \textit{bond length}, \textit{bond angle} and \textit{dihedral angel}. By reconstructing these local descriptors that can fully describe the 3D conformer, the encoder could implicitly generate and encode 3D information in its latent vectors, which can better reflect certain molecular properties \cite{zhu2022unified}.
Considering the weight distribution problem of these three pre-training tasks, we design a pre-training surrogate metric to dynamically adjust the weights of each generative task, and conduct the pre-training of the model as a bi-level optimization problem:
\begin{equation}
\label{equation:bi-level}
\begin{aligned}
\min_{\lambda_{1},\cdots,\lambda_{n}} \mathcal{H}(f_{\theta^{*}}(G)), \quad
\text{s.t.} \, \theta^{*}=
\text{arg}\min_{\theta}\sum_{i=1}^{n}\lambda_{i} \mathcal{L}_{i}(f_{\theta}(G)), 
\end{aligned}
\end{equation}
where $\lambda_{i}$ is the loss weight for pre-training task $\mathcal{L}_{i}$.
As a pre-training surrogate metric, $\mathcal{H}$ is used as a medium to correlate conformer generation with downstream property prediction tasks, thus filling the gap between self-supervised pre-training tasks and downstream supervised tasks, so that the geometric priors can be better generalized to downstream tasks with different supervision signals. The overall framework of 3D PGT is shown in Figure~\ref{fig:framework}.

\subsubsection{3D generative pre-training tasks.}\

% Why Generative pre-training task? [Why length, angle, dihedral angle?]

Generative pre-training task enables the model to understand the DFT based geometry optimization process of molecules from 2D topology to 3D geometry. 
Since molecular geometry is determined by the quantum mechanical behavior of the electrons, 
the generative task indirectly learns the prediction of quantum chemical properties by learning the generation of 3D conformers. 
In this work, we design multiple generative 3D pre-training tasks based on molecular structure generation, including the prediction of bond length, bond angle, and dihedral angles, which are complementary local geometric parameters that can elucidate a 3D conformer. 

As the distance between the equilibrium nuclei of two bonded atoms, bond length is the basic configuration parameter for understanding molecular structure \cite{lide2004crc}. 
We first take bond length prediction as one of the generative pre-training tasks and define the loss function as follow: 
\begin{align}
\mathcal{L}_{\text{length}}({E})=\frac{1}{ \left| {E}  \right|} \sum_{(i,j)\in {E}}(f_{\text{length}}(\textbf{h}_{i}^{(L)},\textbf{h}_{j}^{(L)})-l_{ij})^2,
\end{align}
where, $\textbf{h}_{i}^{(L)}$ and $\textbf{h}_{j}^{(L)}$ denotes the representation of atom $i$ and $j$ which are encoded from $f_{\theta}(\cdot)$, $f_{\text{length}} (\cdot)$ is a network that predicts the bond length, $l_{ij}$ denotes the length of the bond $(i,j) \in {E}$. 
At the same time, the bond angle and dihedral angle which can be understood as the local spatial description of molecular geometry are the complementary information of bond length. These two prediction tasks could be designed as:
\begin{align}
&\mathcal{L}_{\text{angle}}({A}) =\frac{1}{{ \left| A  \right|}} \sum_{(i,j,k)\in {A}} ( f_{\text{angle}}(\mathbf{h}_{i}^{(L)},\mathbf{h}_{j}^{(L)},\mathbf{h}_{k}^{(L)})-\alpha_{j,i,k})^2,\\
&\mathcal{L}_{\text{dihedral}}({D}) = \nonumber \\
& \frac{1}{{ \left| D  \right|}} \sum_{(i,j,k,m)\in {D}} ( f_{\text{dihedral}}(\mathbf{h}_{i}^{(L)},\mathbf{h}_{j}^{(L)},\mathbf{h}_{k}^{(L)},\mathbf{h}_{m}^{(L)})-\phi_{k,i,j,m})^2,
\end{align}
where $A$ denotes the set of bond angles and $D$ denotes the set of dihedral angles. 
$\alpha_{i,j,k}$ and $\phi_{k,i,j,m}$ represent the bond angle formed by node $i,j,k$ and the dihedral angle involving node $i,j,k,m$. Their value ranges are $[0, \pi]$ and $[0, 2\pi]$.
$f_{\text{angle}}$ and $f_{\text{dihedral}}$ are two prediction networks

However, with the number of neighbors $\left| \mathcal{N} \right|$, the $\mathcal{O}({\left| {\mathcal{N}}\right |}^2)$ and $\mathcal{O}({\left| {\mathcal{N}}\right |}^3)$ computational complexity of bond angle calculation and dihedral angle calculation will cause a large amount of time and memory consumption during pre-training. Inspired by \cite{schutt2021equivariant}, we propose a task variant that directly predicts the sum of cosine values of bond angle and dihedral angle as follows: 
\begin{align}
&\mathcal{L}_{\text{angle}}(\mathbf{A})=\frac{1}{\mathbf{ \left| A  \right|}} \sum_{i\in \mathbf{V}}(\sum_{(j,k)\in \mathbf{A}_{i}} f_{\text{angle}}(\mathbf{h}_{i}^{(L)},\mathbf{h}_{j}^{(L)},\mathbf{h}_{k}^{(L)})-\mathbf{\Theta}_{i})^2,\label{eq:angle}\\
&\mathcal{L}_{\text{dihedral}}(\mathbf{D})= \nonumber \\
&\frac{1}{\mathbf{ \left| D  \right|}}\sum_{(i,j)\in \mathbf{E}} (\sum_{(k,m)\in \mathbf{D}_{ij}} f_{\text{dihedral}}(\mathbf{h}_{i}^{(L)},\mathbf{h}_{j}^{(L)},\mathbf{h}_{k}^{(L)},\mathbf{h}_{m}^{(L)})-\mathbf{\Phi}_{(i,j)})^2,
\label{eq:dihedral}
\end{align}
where $\mathbf{A}_{i}$ represents the angle set formed by node $i$, $\mathbf{D}_{ij}$ represents the dihedral angle set with $e_{ij}$ as the common rotation axis. 
The calculation details of $\mathbf{\Theta}_{i}$ and $\mathbf{\Phi}_{(i,j)}$ which denotes the sum of cosine values of $\mathbf{A}_{i}$ and $\mathbf{D}_{ij}$ are described in Appendix \ref{appendix:acceleration}.

\subsubsection{Fusion supervised by a surrogate Metric} \
%\huan{How about a title ``Adaptive fusion by Total Energy'' or ``Task Weight Search by Total Energy''?} \xu{'Adaptive fusion by quality measure' ?}

% Different pre-training tasks will have different effects on downstream tasks. 
In order to integrate the multiple generative tasks we defined, we need a supervision signal to optimize the weights of each pre-training task. 
Given access to the downstream task labels, we can explicitly use the performance on the downstream task as an objective function to guide weight tuning. 
However, considering the variety of molecular property prediction tasks, searching $\lambda_{i}$ as a hyper-parameter can be highly expensive. Therefore, designing a generalized surrogate metric for searching these weights is necessary.

The total energy of a molecule is represented by the sum of the energy of each atomic domain and the energy of electrostatic interactions between the domains \cite{kim2019novel}.
As mentioned above, the geometry optimization of DFT is to search the local minima on the potential energy surface \cite{axelrod2022geom}, as shown in Figure~\ref{fig:generative}.
The basic functional form of potential energy in molecular mechanics includes the bonded terms for interactions of atoms and the nonbonded terms that describe the long-range electrostatic and van der Waals forces, which means that the \textit{total energy} of a molecule can reflect its geometric information. 
Considering that the \textit{total energy} $E_{\text{total}}$ as an attribute of 3D conformer can be acquired along with conformer generation, we design a graph-level prediction task based on the  $E_{\text{total}}$, which can be expressed as: 
\begin{equation}
\mathcal{L}_{E_{\text{total}}} (f_{\theta}(G)) = (f_{\theta}(G) - E_{\text{total}})^2. 
\end{equation}
Here, we use $\mathcal{L}_{E_{\text{total}}}$ as the surrogate metric to optimize $\lambda_{i}$, and due to the association between 3D geometric information and molecular properties, $\mathcal{L}_{E_{\text{total}}} (G)$ can be used to supervise the fusion of these generative pre-training tasks so that the geometric prior obtained by pre-training is also generalized for downstream property prediction tasks. 

However, for the bi-level optimization problem in equation (\ref{equation:bi-level}), 
computing the gradient of $\mathcal{L}_{E_{\text{total}}}$ to update $\lambda_{i}$ 
is very expensive~\cite{pedregosa2016hyperparameter}.
%\footnote{+qm+ gradient can be computed,
%	the problem is that the computation process needs $\theta^*$ and Hessian,
%	which are very expensive.
%	see (2) at 
%	\url{https://arxiv.org/pdf/1602.02355.pdf} \xu{\checkmark}} 
Inspired by \cite{finn2017model}, 
we use mate-gradients to optimize this bi-level problem. 
By unrolling the training procedure, 
the meta-gradient $\nabla^{\text{meta}}_{\{\lambda_{i}\}}$ can  be expressed as: 
\begin{align}
\nabla^{\text{meta}}_{\{\lambda_{i}\}} :&= \nabla_{\{\lambda_{i}\}} \mathcal{L}_{E_{\text{total}}}(f_{\theta_{T}}(G)) \\
&= \nabla_{f_{\theta_{T}}} \mathcal{L}_{E_{\text{total}}}(f_{\theta_{T}}(G)) \cdot \nabla_{\theta_{T}} f_{\theta_{T}} (G) \nabla_{\{\lambda_{i}\}} \theta_{T},
\end{align}
where $\theta_{T}$ denotes the encoder's parameters $\theta$ at step $T$, which is depends on the task weights $\{\lambda_{i}\}$. Thus, the gradient descent performed on $\{\lambda_{i}\}$ can be achieved by meta-gradient $\nabla^{\text{meta}}_{\{\lambda_{i}\}}$ as $\{\lambda_{i}\} -\eta \nabla^{\text{meta}}_{\{\lambda_{i}\}}$, where $\eta$ is the learning rate for outer optimization. At the same time, inspired by DARTS \cite{liu2018darts}, we only perform one step gradient descent on $\theta$ and update $\{\lambda_{i}\}$ each iteration, which means $\{\lambda_{i}\}$ is dynamic during the pre-training stage and adjusts between $[0, 1]$.
The calculation details of $\nabla^{\text{meta}}_{\{\lambda_{i}\}}$ can be referred to \cite{finn2017model}.

\begin{table*}[t]
	%\small
	\caption{Results for quantum properties prediction in QM9. The evaluation metric is MAE (Smaller is better). The compared methods include 2D models without pre-training, pre-training baselines, and the 3D model SMP. \textit{True 3D} means to input the ground truth 3D structure, and \textit{RDkit} means to input the structure calculated by RDkit. The best performance of each target is marked in bold. }
	\label{tab:qm9}
	\begin{tabular}{l|cc|ccccccc|cc}
		\toprule
		\multirow{2}{*}{Target}   & \multicolumn{2}{c}{2D Model} & \multicolumn{7}{c}{Pre-training Baselines}  & \multicolumn{2}{c}{3D SMP} \\
		& PNA & GPS & GraphCL & AttrMask & GPT-GNN &  GraphMVP  & 3D Infomax &  $E_{\text{Total}}$ & 3D PGT & \textit{RDKIT} & \textit{True 3D} \\
		\midrule
		$\mu$         &  0.4133 & 0.4087 & 0.3937  & 0.4626 & 0.3975  & 0.3489  & 0.3507  & 0.3927 & \textbf{0.3409} & 0.4344  & 0.0726 \\
		$\alpha$  &  0.3972 & 0.3884 & 0.3295  & 0.3570 & 0.3732  & 0.3227  & 0.3268   & 0.3514 & 0.3121 & \textbf{0.3020}  & 0.1542 \\
		$HOMO$    &  82.10  & 81.19     & 79.57  & 80.58    & 93.11    & 68.62  & 68.96  & 78.28  & \textbf{68.24}  & 82.51    & 56.19  \\
		$LUMO$   &  85.72 & 84.86   & 80.81  & 84.93     & 99.84   & 70.23 & \textbf{69.51}  & 81.26  & 69.73 & 80.36   & 43.58  \\
		GAP           &  123.08 & 121.17  & 120.08   & 116.21  & 131.99   & 101.84  & 101.71   & 112.37  & \textbf{101.53} & 114.24  & 85.10  \\
		R2            &  22.14  & 21.78   & 21.84  & 29.23    & 29.21    & 17.03  & 17.39  & 21.86  & \textbf{16.89}  & 22.63   & 1.510  \\
		ZPVE      &  15.08  & 12.29   & 12.39  & 25.91    & 11.17      & 7.958  & 7.960     & 18.28  & 7.924 &  \textbf{5.180}      & 2.690  \\
		$c_{v}$   &  0.1670 & 0.1618 & 0.1422  & 0.1587 & 0.1795   & 0.1287  & 0.1306  & 0.1507 & \textbf{0.1217} & 0.1419   & 0.0498  \\
		\bottomrule 
	\end{tabular} 
\end{table*}

\subsection{Further Improvements}

\subsubsection{Pre-training by Multiple conformers} \

As mentioned in Section~\ref{section:3.1}, 3D molecular datasets usually calculate the conformer with the lowest energy. However, leveraging structural information of multiple conformers could bring additional benefits \cite{stark20223d, liu2022pre}. Therefore, we consider introduce multiple conformers $\{\mathbf{P}_{i}^{c}\}_{c\in\{1...C\}}$ of the same molecule graph $G_{i}$ in the pre-training datasets, and repeat the lowest energy conformer if the conformer number of a molecule is fewer than $C$. 
For the 3D pre-training in our framework, we here simply consider it as a data augmentation and avoid the pipeline changes of pre-training. 

\subsubsection{Architecture search of Graph Transformer} \ 

\label{tab:arch search}
In addition to using the surrogate metric $\mathcal{H}$ to search the weight of each pre-training task, we also search the architecture of pre-training backbone. 
Based on the hybrid Graph Transformer structures which combined the local message passing module and global attention module, we build a search space that includes the above modules to search the backbones for different pre-training datasets. 
Specifically, in each layer of the graph encoder $f_{\theta}(\cdot)$, the search space of operations $\mathcal{O}_{GT}$ includes 3 operations: local message passing, global attention and a hybrid operation through a Feed Forward Network (FFN). We then relax the discrete selection of operation by a weighted summation of all possible operations, and use a differentiable search algorithm based on Gumble-Softmax \cite{jang2016categorical}. 
The details of the architecture search are described in Appendix \ref{method_details}.

\section{Experiments}

In this section, we conduct various experiments to demonstrate the effectiveness of the proposed 3D PGT. As explained in Introduction, we aim to address the scenario where only 2D molecular graphs are available for prediction, thus we can avoid the huge computational cost of generating 3D conformer for each molecule. Thus in the experiments, we pre-train using subsets of 3D molecular structure datasets, and then fine-tune the models in downstream tasks of 2D datasets or 3D datasets while ignoring the 3D structures.

Here we briefly explain the organization of our experiments. Firstly, we evaluate the proposed 3D PGT on popular benchmark dataset, \textbf{QM9} \cite{ramakrishnan2014quantum} , for quantum chemistry properties. Then broader downstream applications, including binary property classification, regression property, and drug-target affinity prediction, are designed to evaluate the generalization/transfer ability of the 3D PGT. Further ablation studies are conduct to show the importance of each design modules of our method. Finally, more interestingly, we apply 3D PGT to property prediction in a large-scale molecular dataset, \textbf{PCQM4Mv2} \cite{nakata2017pubchemqc}, from the famous large-scale open graph benchmark (OGB) \cite{hu2021ogb}, where 3D structures are only given in training dataset, to justify the practical values of the problem we try to address in this work, as well as the technical contributions.

Considering the different settings of the designed experiments, we give the introduction of experimental setups, including datasets, evaluation metrics, and baselines, in the corresponding subsections. And more detailed information are given in Appendix \ref{append:data} and \ref{append:baseline}.

\noindent\textbf{Implementation of 3D PGT.}
%\huan{briefly introduce the key info of our model, and refer to appendix for more information. Besides, give the code link here.} \xu{\checkmark.}
We choose GPS \cite{rampavsek2022recipe} as the Graph Transformer for pre-training due to its state-of-the-art performance on OGB-LSC \cite{hu2021ogb}. As mentioned in Section~\ref{tab:arch search}, we further search the architecture by the surrogate metric $\mathcal{L}_{E_{\text{total}}}$, the details are described in Appendix \ref{tab:implement details}. To implement 3D PGT and reproduce the results, the code is provided as an anonymous link \footnote{\url{https://github.com/LARS-research/3D-PGT}}. 

%In the experimental part, considering the cost of explicitly computing 3D conformations in downstream tasks, we focus on organizing pre-training on available 3D molecular datasets and evaluating the improvement of transferring to downstream tasks where 3D information is not available. The 3D datasets we selected for pre-training include: \textbf{QM9} \cite{ramakrishnan2014quantum} with 134k small moleucles, \textbf{GEOM-Drugs} \cite{axelrod2022geom} with 304k molecules and \textbf{PCQM4Mv2} \cite{nakata2017pubchemqc} with 3.74 million. 
%To evaluate the benefits of pre-training, we then finetune the models and predict properties on the subsets of QM9 while masking the 3D structure. We also report the results on non-quantum properties of 8 binary molecular property prediction tasks, 4 regression tasks and 2 drag-target affinity tasks to demonstrate the generalizability of our pre-trained framework. Finally, we simulate the scenario when 3D geometric information is not available on a large-scale dataset. More detailed explanation are in Appendix.

\begin{table*}
	%\small
	\caption{Results for eight molecular property prediction tasks (classification). The evaluation metric is ROC-AUC (Larger is better). The best performance for each task is marked in bold.}
	\label{tab:classification}
	\begin{tabular}{lcccccccccc}
		\toprule
		Method & Pre-training & BBBP & Tox21 & ToxCast & Sider & ClinTox & MUV & HIV & Bace & Avg.\\
		\midrule
		GIN & \XSolidBrush & 65.4 & 74.9 & 61.6 & 58.0 & 58.8 & 71.0 & 75.3 & 72.6 & 67.21 \\ 
		GraphGPS & \XSolidBrush & 67.0 & 71.5 & 68.5 & 56.4 & 71.1 & 66.9 & 77.0 & 76.9 & 69.41 \\
		\midrule
		AttrMask & \Checkmark & 69.4 & 73.8 & 61.9 & \textbf{60.6} & 68.3 & 73.5 & 73.4 & 76.8 & 69.71 \\
		ContextPred & \Checkmark & 71.1 & 73.1 & 62.6 & 59.1 & 73.8 & 73.1 & 72.6 & 78.7 & 70.60 \\
		DistancePred & \Checkmark & 64.7 & 74.3 & 60.6 & 56.5 & 55.7 & 73.2 & 75.7 & 65.1 & 65.84 \\
		EnergyPred & \Checkmark  & 66.8  &  75.6 & 61.2  &  58.7  & 56.4  & 72.7  & 74.4  & 65.5 & 66.41 \\
		GraphMVP & \Checkmark & 70.8 & \textbf{75.9} & 63.1 & 60.2 & 79.1 & \textbf{77.8} & 76.2 & 79.3 & 72.79 \\ 
		3D Infomax & \Checkmark & 69.1 & 74.9 & 64.4 & 53.4 & 59.4 & 73.1 & 76.8 & 79.4 & 68.69 \\
		\midrule
		\textbf{3D PGT}  & \Checkmark & \textbf{72.1} & 73.8 & \textbf{69.2} & \textbf{60.6} & \textbf{79.4} & 69.4 & \textbf{78.1} & \textbf{80.9} & \textbf{72.94} \\ 
		% \textbf{Our method*} & \Checkmark \\ 
		\bottomrule
	\end{tabular}
\end{table*}

\noindent\textbf{Datasets.} 
We first explore the performance on eight quantum properties prediction of QM9 \cite{ramakrishnan2014quantum}, which contains 134k molecules with a single conformer. Following the experimental setup of mainstream 3D pre-training work \cite{stark20223d}, we randomly select 50k molecules with geometric information from QM9 for pre-training. Then, we pick another 50k molecules from the rest and mask their 3D structures for model fine-tuning. We use the scaffold splits with an 8/1/1 ratio and report the Mean Absolute Error (MAE) of 8 quantum chemical properties in the test dataset.

\noindent\textbf{Baselines.} 
As the SOTA performance on molecular tasks, we first select PNA \cite{corso2020principal} as the 2D model for comparison. It also includes GPS \cite{rampavsek2022recipe}, the Graph Transformer that achieved SOTA on the OGB challenge \cite{hu2021ogb} which is also the backbone of our 3D PGT. At the same time, we compare the most well-known pre-training methods, including AttrMask \cite{hu2019strategies}, GPT-GNN \cite{hu2020gpt}, GraphCL \cite{axelrod2020molecular}, GraphMVP \cite{liu2022pre} and 3D Infomax \cite{stark20223d}. The backbone and hyperparameter settings of the above pre-training methods follow their original settings. 
We design a baseline that directly uses the \textit{total energy} prediction as a pre-training task, so as to prove the significance of using \textit{total energy} as a surrogate metric by comparison. 
We also chose the SMP \cite{zitnick2022spherical} based on the 3D Graph Network framework as a 3D model for comparison. The configuration of SMP includes the real 3D structure calculated by DFT from the QM9 dataset; and the relatively rough 3D structure calculated based on RDKit \cite{landrum2016rdkit}.

\begin{figure}[t]
	\centering
	\includegraphics[width=0.85\linewidth]{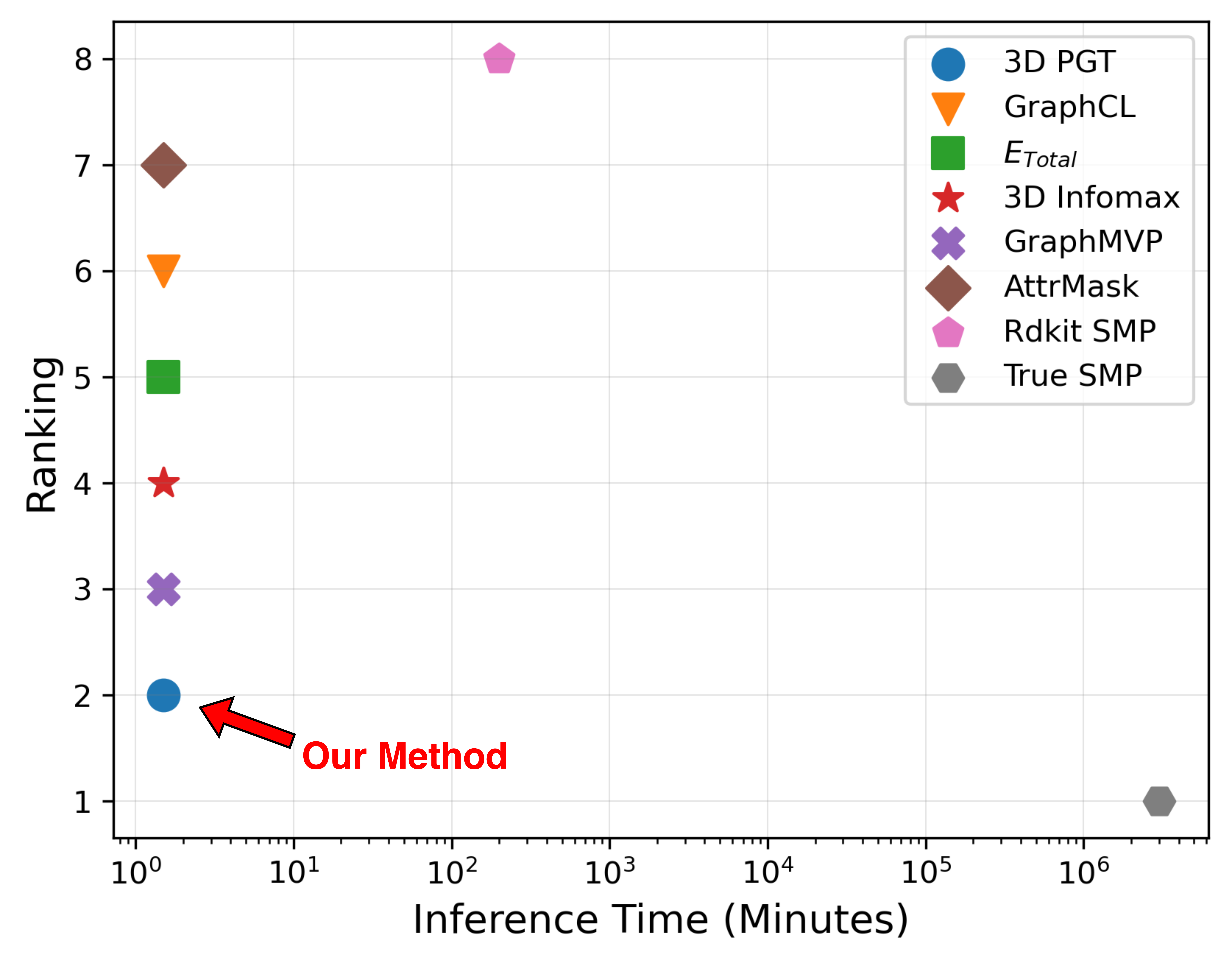}
	\caption{The total inference time and performance ranking of different methods for $HOMO$ prediction.}
	\label{tab:efficiency}
	\vspace{-10pt}
\end{figure}
\subsection{Quantum Chemistry Properties Prediction}

\noindent\textbf{Results.}
The compared results are shown in Table \ref{tab:qm9}. 
It can be seen that 3D PGT achieves better or competitive performance compared to all baselines. 
First, we can find that our 3D pre-training framework brings an average 17.7\% MAE reduction compared with the GPS backbone, while significantly outperforming the 2D pre-training methods. It is obvious that the improvement brought by the pre-training without geometry in GraphCL is limited, and the same limitation is also reflected in the pre-training which uses energy as the only optimization objective.
Besides, the latest two pre-training models on molecular 3D structures, i.e., GraphMVP and 3D Informax, achieve evident performance improvement compared others only pre-training on 2D molecular graphs, which verifies the usefulness of 3D geometry in molecular prediction. Then the further improvement of our 3D PGT compared GraphMVP and 3D Infomax demonstrates the effectiveness of our approach. 
It is clear that SMP achieves the best performance when given the 3D structures in the test dataset, while notably that our proposed 3D PGT beat SMP with coarse 3D information in 6 targets out of 8, which indicates the advantage of this 3D pre-training and fine-tuning paradigm.

In order to explicitly analyse the efficiency comparison of 2D, 3D models and pre-trained models, we statistics the total inference time and performance ranking of all baselines on the test set to analyze the efficiency between different methods, the result of \textit{HOMO} prediction is shown in Figure~\ref{tab:efficiency}. 
It can be seen that the 3D Pre-training based methods achieve a better trade-off between inference time and performance, especially 3D PGT. Although SMP using explicit ground truth 3D information has obvious advantages in predicting molecular properties, when the rough 3D information calculated by RDKIT is input, its performance is lower than that of 3D pre-training methods, which means that it is difficult for 3D models to balance computational efficiency and performance.

\begin{table*}
	%\small
	\caption{Results for four molecular property prediction tasks (regression) and two drug-target affinity tasks (regression). The evaluation metrics are RMSE and MSE, respectively (Smaller is better for both). The best performance for each task is marked in bold.}
	\label{tab:regression}
	\begin{tabular}{lcccccc|ccc}
		\toprule
		\multirow{2}{*}{Method} &  \multirow{2}{*}{Pre-training} & \multicolumn{5}{c}{Molecular Property Prediction (\textbf{RMSE})} &  \multicolumn{3}{c}{Drug-Target Affinity (\textbf{MSE})} \\
		\cmidrule{3-10}
		&  & ESOL & Lipo & Malaria & CEP & Avg & Davis & KIBA & Avg.  \\
		\midrule
		GIN & \XSolidBrush & 1.178  & 0.744 & 1.127 & 1.254 & 1.0756 & 0.286 & 0.206 & 0.2459  \\ 
		GraphGPS & \XSolidBrush & 1.097 & 0.721 & 1.117 & 1.239 & 1.0435 & 0.284 & 0.203 & 0.2435 \\
		\midrule
		AttrMask & \Checkmark & 1.118 & 0.733 & 1.121 & 1.257 & 1.0573 & 0.293 & 0.205 & 0.2491 \\
		ContextPred & \Checkmark & 1.196 & 0.702 & \textbf{1.101} & 1.243 & 1.0606 & 0.279 & 0.198 & 0.2382  \\
		DistancePred & \Checkmark & 1.126 & 0.718 & 1.157 & 1.291 & 1.0714 & 0.282 & 0.193 & 0.2375  \\
		EnergyPred & \Checkmark & 1.118 & 0.711 & 1.126 & 1.277 & 1.0595 & 0.291 & 0.185 & 0.2380  \\
		GraphMVP & \Checkmark & 1.072 & 0.701 & 1.107 & 1.231 & 1.0278 & 0.277 & 0.177 & 0.2270  \\ 
		3D Infomax & \Checkmark & 1.087 & 0.706 & 1.121 & 1.218 & 1.0371 & 0.278 & 0.176 & 0.2301  \\
		% 3D Infomax & \Checkmark & 69.1 & 74.5 & 64.4 & 53.4 & 59.4 &  & 76.1 & 79.4  \\
		\midrule
		\textbf{3D PGT}  & \Checkmark & \textbf{1.061} & \textbf{0.687} & 1.104 & \textbf{1.215} & \textbf{1.0168} & \textbf{0.275} & \textbf{0.173} & \textbf{0.2240} \\ 
		% \textbf{Our method*} & \Checkmark \\ 
		\bottomrule
	\end{tabular}
\end{table*}

\subsection{Generalizing to different downstream tasks}

%\huan{How about a title like ``Generalize to different downstream tasks''? \xu{\checkmark}}
\noindent\textbf{Datasets.}
In addition to predict the quantum chemical properties which is closely related to molecular geometry, we further generalize the downstream tasks to pharmacology, physical chemistry and biophysics, which only measured 2D structures of molecules.
Following the experimental settings in \cite{liu2022pre}, we randomly select 50k molecules in GEOM \cite{axelrod2022geom} with single conformer for pre-training, and finetune on 8 mainstream binary classification tasks and 6 classification tasks. These downstream datasets are all in the low-data regime and we describe their details in Appendix.

\noindent\textbf{Baselines.}
Following the baseline chosen by \cite{liu2022pre}, in addition to the well-performing backbone GIN \cite{xu2018powerful} and GPS \cite{rampavsek2022recipe}, we also selected two 2D pre-training baselines, AttrMask \cite{hu2019strategies} and ContextPred \cite{hu2019strategies};  two predictive pre-training tasks DistancePred \cite{zhu2022unified} and EnergyPred and the latest SOTA solution GraphMVP \cite{liu2022pre} and 3D Infomax \cite{stark20223d}. 

\noindent\textbf{Results of MoleculeNet Benchmark.}
MoleculeNet is a popular benchmark for molecular property prediction. We first compare 3D PGT with other baselines on eight binary property prediction tasks.
The performance is shown in Table \ref{tab:classification}, where the best results are marked in bold. It can be seen that 3D PGT outperforms other baselines on most of these downstream tasks, which means our method can generalize the benefits of our pre-training framework to a broader range of downstream tasks, even though most of these tasks have only a small number of samples. 
We observe that 3D PGT does not achieve the best performance on Tox21 and MUV tasks, which is mainly because the GPS itself as the backbone is easy to overfit on these datasets. However, it is certain that compared to GPS, 3D PGT can still achieve stable performance gains through pre-training.

\noindent\textbf{Results of regression and drug-target affinity tasks.}
To further confirm whether 3D pre-training can help in a wider range of downstream tasks, we organized additional experiments on 4 additional molecular property regression tasks and 2 Drug-Target Affinity (DTA) tasks. 
Different from the prediction of the molecular properties themselves, the DTA task is to predict the affinity score between the molecular drug and the target protein. 
As shown in Table \ref{tab:regression}, the results show that 3D PGT is still competitive in these 6 regression tasks, and it also proves that 3D Pre-training can help in DTA tasks.

%\subsection{Transfer from different pre-training datasets}
%It is proved that our pre-training method does not have negative transfer phenomenon, and can obtain benefits stably. This is due to:
%\begin{enumerate}
%  \item  Compared with 2D pre-training, 3D pre-training can help the model better understand the molecular structure, not only the topological relationship, but also more information; 
%  \item Furthermore, our auto-method can allocate the most appropriate weight, so as to obtain the most stable benefits on downstream tasks. 
%  \item Finally, it is proved that our pre-training method can span samples belonging to different chemical spaces and has stronger generalization ability. Therefore, we can use a wider range of datasets for pre-training and finetune more tasks. 
%\end{enumerate}

\begin{figure*}[t]
	\centering
	\includegraphics[width=0.86\linewidth]{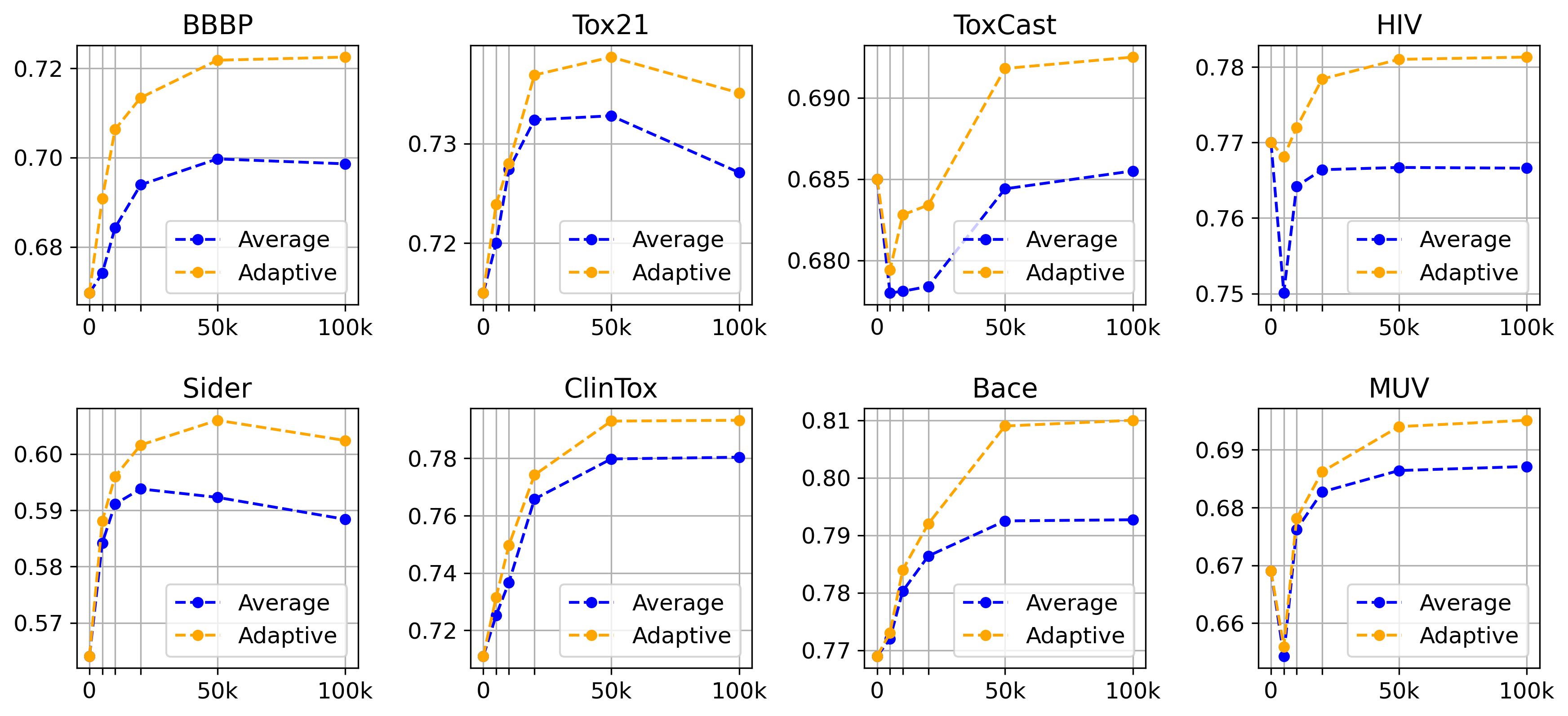}
	\caption{The AUC of 8 classification tasks when using different numbers of molecules during pre-training. The blue broken line represents the use of the same weight for all pre-training tasks, and the yellow represents using the searched adaptive weights.} 
	\label{tab:datasize_8class}
\end{figure*}

\subsection{Ablation studies}
\label{sec:exp:aba}
%
%\subsubsection{Impact of each module.} \
%
%First,We conduct an ablation study to verify the contribution of each module, as shown in Table \ref{tab:ablation pre-training}. 

\begin{figure}[htbp]
	\centering
	\includegraphics[width=0.7\linewidth]{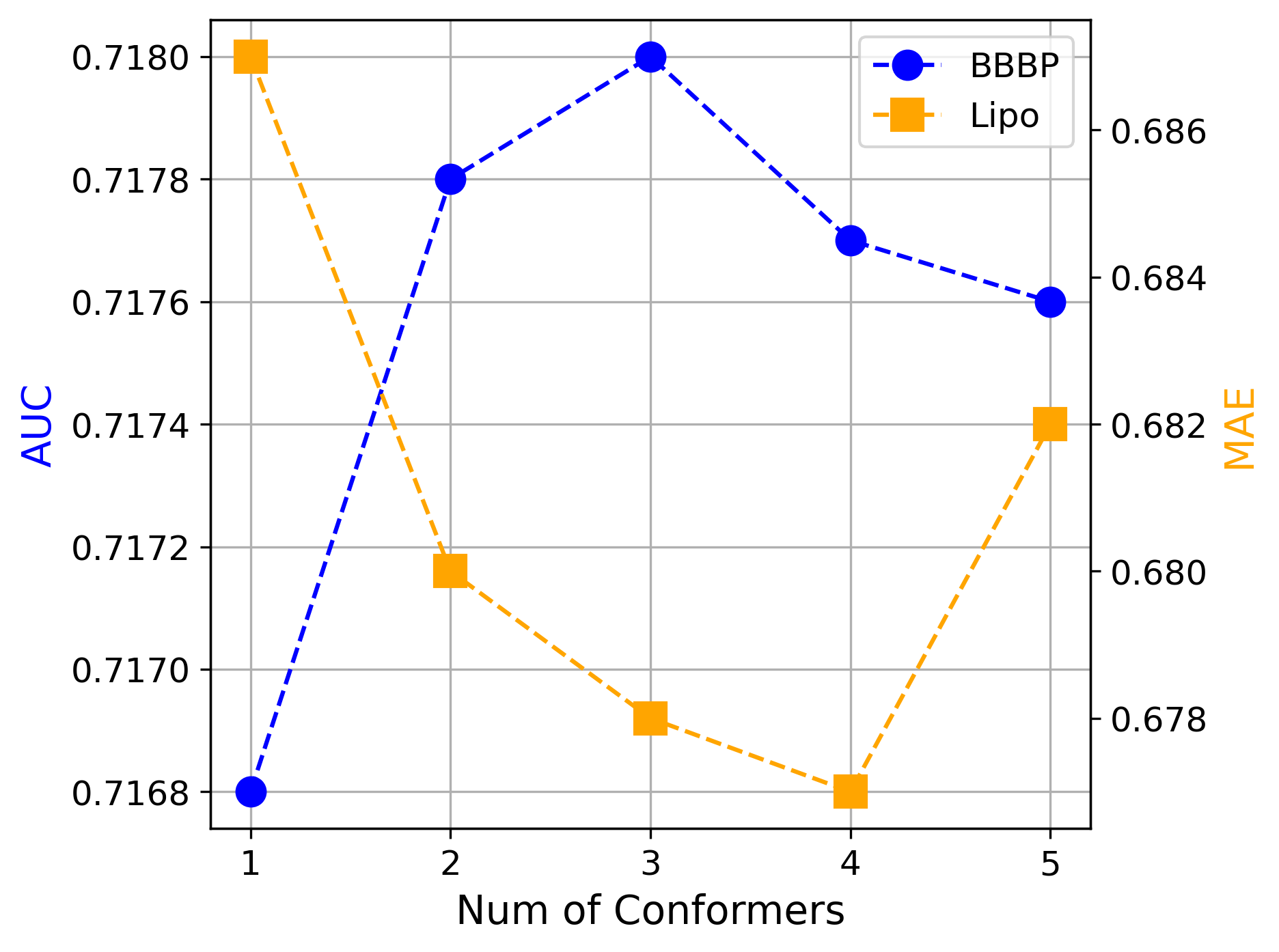}
	\caption{The performance when varying the number of conformers used during pre-training.}
	\label{tab:conformers}
\end{figure}

\begin{figure}[htbp]
	\centering
	\includegraphics[width=0.92\linewidth]{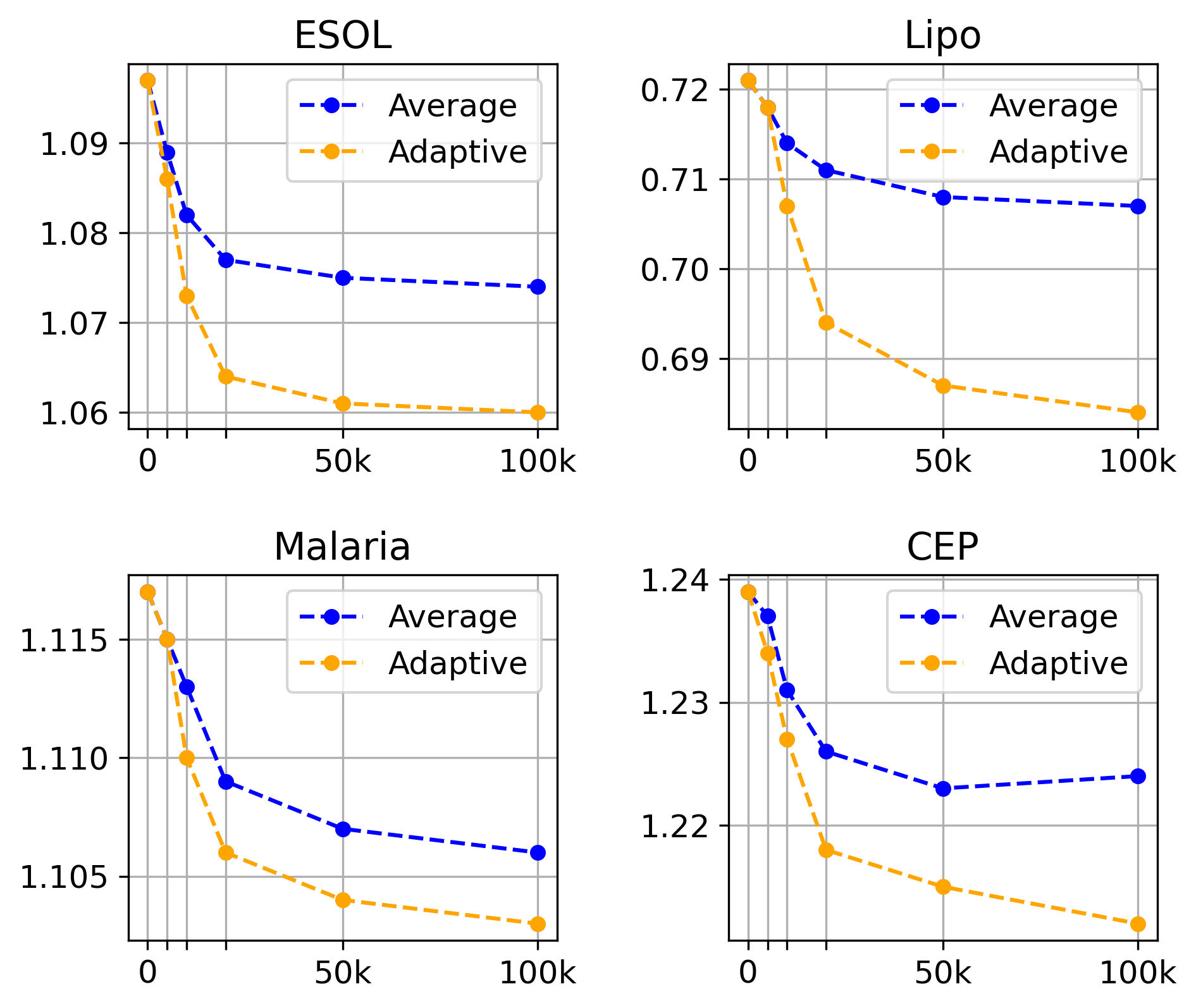}
	\caption{The RMSE of 4 regression tasks when varying the number of pre-training molecules. The blue broken line represents the use of the same weight for all pre-training tasks, and the yellow represents using the searched adaptive weights.}
	\label{tab:datasize_6regreesion}
	\vspace{-10pt}
\end{figure}

\noindent\textbf{The effect of searched adaptive weight for multiple tasks.} 
In Section~\ref{tab:ourmethod}, we propose a surrogate metric based on total energy to search the adaptive weight of each generative pre-training task. 
To verify its effectiveness, we design an ablation study by pre-training a variant of average pre-training. 
We compared the two methods of average training and the weight search, 
the results in Figure~\ref{tab:datasize_8class} demonstrate that the searched weights lead to greater benefits on various downstream tasks. 
And the results of the six regression tasks also reached the same conclusion, as shown in Figure~\ref{tab:datasize_6regreesion}. 

\noindent\textbf{The influence of pre-training dataset sizes.} 
From Figure~5, it can be seen that for most downstream tasks, the performance increases with the amount of pre-trained data, although there is a marginal effect. 
However, on Tox21 and Sider, when the pre-training dataset is further increased, the benefits of pre-training will decrease. One possible reason is that the backbone of GPS itself exhibits overfitting on the two data sets, and further injection of prior knowledge would not alleviate it \cite{sun2022does}.

\noindent\textbf{Number of conformers for pre-training.} 
For the multiple conformers of a single molecule, we verify whether useing these conformers can benefit the downstream task performance, 
as shown in Figure~\ref{tab:conformers}. The experiment shows that adding conformers for pre-training within a certain range can bring further pre-training benefits. 
However, consistent with the findings of previous work \cite{stark20223d, liu2022pre, axelrod2020molecular}, we observe that there is a bottleneck in the improvement brought by adding more conformers. 
One accepted conclusion is that the top-5 conformers sampled according to the energy are sufficient to cover 80\% of the equilibrium state, and further addition of conformers cannot supplement more abundant geometric prior for downstream tasks.
As discussed in Section~\ref{sec:exp:aba}, 
we observe that the GPS backbone is prone to overfitting on these two few-shot datasets, 
and the results show that further introducing too many geometric priors may not lead to better gains in this case.

%\subsection{Large-scale Pre-training on PCQM4Mv2}
\vspace{-2pt}
\subsection{Pre-training on Large-scale Dataset}

  \begin{table}[t]
		\small
	\caption{Results on PCQM4Mv2 validation set in OGB Large-Scale Challenge. The evaluation metric is the Mean Absolute Error (MAE) [eV] (smaller is better). We report and compare with single model from the top solutions on OGB Leaderboard. The best performance for each task is marked in bold.}
	\label{tab:pcqm4mv2}
		\vspace{-5pt}
	\begin{tabular}{clcccccc}
		\toprule
		& Method & 3D Pre-training & \#Param. & Valid MAE \\
		\midrule
		%& GCN & \XSolidBrush  & 2.0 M & 0.1379 \\
		& GIN   & \XSolidBrush  & 3.8 M & 0.1195 \\
		& GRPE  & \XSolidBrush  & 46.2 M & 0.0890 \\
		& EGT  & \XSolidBrush  & 89.3 M & 0.0869 \\
		& Graphormer  & \XSolidBrush  & 48.3 M & 0.0864 \\
		& TokenGT   & \XSolidBrush  & 48.5 M & 0.0910 \\
		& GPS$_{\text{medium}}$   & \XSolidBrush & 19.4 M & 0.0858 \\
		& GPS$_{\text{deep}}$   & \XSolidBrush  & 138.1 M & 0.0852 \\
		\midrule
		& GEM-2  & \Checkmark  & 32.1 M & 0.0793 \\ 
		& Global-ViSNet  & \Checkmark & 78.5 M & 0.0784 \\
		& Transformer-M  & \Checkmark  & 47.1 M & 0.0787 \\
		& GPS++ & \Checkmark & 44.3M & 0.0778 \\
		\midrule
		% & {\textbf{3D PGT}}  & \Checkmark & 44.7M & 0.0764 \\ 
		& \textbf{3D PGT}  & \Checkmark & 42.6M & \textbf{0.0762} \\ 
		\bottomrule
	\end{tabular}
%\vspace{-10pt}
\end{table}

\noindent\textbf{Datasets.} As the graph-level track of OGB-LSC \cite{hu2021ogb}, PCQM4Mv2 is a quantum chemistry dataset under the PubChemQC project \cite{nakata2017pubchemqc}. 
The dataset contains a total of 3.74 million molecules, of which the 3D geometric information of 3.37 million training samples is obtained by DFT optimization.
We adhered to the default train-validation split provided by OGB, and the test set was reserved for the competition and leaderboard and not publicly disclosed.
In order to approach the large-scale virtual screening scenario, the challenge \cite{hu2021ogb} does not provide 3D conformers of the valid dataset and test dataset, and requires the property inference of 150k molecules to be completed within 4 hours using a single GPU, which means that it is not feasible to calculate the geometry of all test samples in the inference stage. Therefore, existing 3D GNN, e.g., SMP, can not work on this challenge due to the expensive computation cost of conformer generation. 

\noindent\textbf{Baselines.}
Because the label of test datasets is officially hidden, we compared the results of validation with the top-tier method on the OGB leaderboard \footnote{\url{https://ogb.stanford.edu/docs/lsc/leaderboards/\#pcqm4mv2}}, which includes GRPE \cite{park2022grpe}, TokenGT \cite{kim2022pure}, EGT \cite{hussain2022global} , GPS \cite{rampavsek2022recipe}, GEM-2 \cite{liu2022gem}, Vis-Net \cite{wang2022visnet} and Transformer-M \cite{luo2022one}. In addition, we also compared GPS++ \cite{masters2022gps++}, the winner solution at OGB LSC @ NeruIPS 2022 Challenge.
%\footnote{\url{https://ogb.stanford.edu/neurips2022/results/\#leaderboard_pcqm4mv2}}.
\textbf{Note} that we only report the performance of single models from these solutions for fair comparisons, though it is clear that various engineering tricks, like ensemble, can be applied to further improve the performance of all methods including our 3D PGT.

\noindent\textbf{Results.} 
The results of validation MAE are shown in Table \ref{tab:pcqm4mv2}.
First, compared with the existing GNNs and Graph Transformers which do not consider 3D geometry information, 3D PGT has been significantly improved by introducing generative pre-training tasks. For GPS, 3D PGT has achieved $10.6\%$ relative MAE reduction by the designed pre-training framework. 
Our 3D PGT outperforms the champion solution GPS++ in terms of single-model performance, proving the advantage of our pre-training framework while using the same backbone. 

\section{Conclusion and Future Work}

In this work, we proposed 3D PGT, a 3D pre-training framework which focus on incorporating 3D information for benefiting the molecular property prediction when the 3D structures is unavailable. 
In 3D PGT, we designed multiple generative pre-training tasks which can bring geometric prior to finetune stage. To better integrate these pre-training tasks and make their benefits generalize, we designed a surrogate metric of pre-training to search the adaptive weight of each pre-task. The experiment we designed proves that the potential geometric prior is not only beneficial to quantum chemical property prediction, but also to the prediction in pharmacology, physical chemistry and biophysics, etc. Moreover, 3D PGT outperforms all baselines from top solutions for large-scale molecular prediction in OGB leaderboard.

For future work, it will be interesting to explore the effectiveness of our proposed framework on larger molecules, e.g. the development of catalysts for storing renewable energy to address global warming \cite{chanussot2021open, tran2022open}, beyond the small molecules in this work.
%Not limited to the performance of 3D PGT, we observed that the increase of pre-training data size has a bottleneck on the benefits of downstream tasks, which encourages us to further explore how to implement pre-training on large-scale 3D molecular datasets efficiently. At the same time, we will also consider introducing backbones compatible with 3D information to play a role in richer downstream scenes, such as molecular simulation and protein folding.

\section*{Acknowledgment}

Q. Yao is supported by NSF of China (No. 92270106).

\clearpage
\bibliographystyle{ACM-Reference-Format}
\balance
\bibliography{sample-base}


\clearpage
\appendix

%\section{Appendix}

\section{Dataset Details}
\label{append:data}

\subsection{Pre-training Datasets with 3D Geometry}

Following the mainstream experimental setup, we use three datasets containing 3D information for pre-training, and the detailed parameters are shown in Table \ref{tab:datasets}. The three datasets are: 
 \begin{itemize}
 \item{\textbf{QM9} \cite{ramakrishnan2014quantum}} contains 134k molecules carrying 3D coordinates, each containing only one low-energy conformer. The largest atoms in QM9 are just nine heavy atoms, and each molecule is annotated with 12 quantum mechanical features.
 \item{\textbf{GEOM-Drugs} \cite{axelrod2022geom}} contains 304k molecules related to biology and pharmacology, and each molecule contains multiple 3D conformers, and Gibbs free energy and integrated energy are annotated as regression targets. The entire dataset contains 16 elements in total. 
 \item{\textbf{PCQM4Mv2}} \cite{hu2021ogb} contains 3.74m molecules with HOMO-LUMO energy gap annotated as regression targets, it is a quantum chemistry dataset originally curated under the PubChemQC project \cite{nakata2017pubchemqc}. A total of 3.37m samples of low-energy conformers in the data set are annotated, while the remaining molecules that only contain 2D structures will be used as the validation set and the test set for the competition.
 \end{itemize}

% \section{Graph Transformer for Molecular Representation Learning}
% \textbf{MPNN for graph-level tasks.} Modern GNN is mostly based on the Message Passing framework \cite{gilmer2017neural} to use a neighborhood aggregation approach to update the representation of the target node. 

\section{Further Method Details}
\label{method_details}
Due to space limitations, the baseline details, technical details, and additional relationship study results compared in the experiment can be seen here: \url{https://github.com/LARS-research/3D-PGT/blob/main/Supplementary_Appendix.pdf}
\subsection{Computational Acceleration of Bond Angles and Dihedral Angles}
Due to the high computational complexity of bond angles and dihedral angles, we reduce the computational complexity of bond angle and dihedral angle to linear time complexity, and the search of bond angle and dihedral angle's node index is replaced by the traversal of the node set ${V}$ and the edge set ${E}$. 
Inspired by the Runtime Geometry Calculation (RGC) \cite{wang2022ensemble} which directly calculates the geometric information from the direction unit which only sums the vectors from the target node to its neighbors once, we design two loss functions based on the sum of cosine values of bond angle and dihedral angle. RGC calculates the sum of bond angle's cosine values as follows:

\begin{align}
&\vec{u}_{ij} =   \frac{ \vec{r}_{ij}} {\Vert { \vec{r}_{ij} } \Vert }, \quad \vec{u}_{ik} =   \frac{ \vec{r}_{ik}} {\Vert { \vec{r}_{ik} } \Vert },\\
&\Theta_{i}=\sum_{j=1}^{N_{i}} \sum_{k=1}^{N_{i}} \cos \alpha_{jik} = \sum_{j=1}^{N_{i}} \sum_{k=1}^{N_{i}} \left\langle \vec{r}_{ij}, \vec{r}_{ik} \right\rangle,
\end{align}
where $\vec{r}_{ij}$ is the vector from node $i$ to its neighboring node $j$. Record $\vec{u}_{ij}$ as the unit vector of $\vec{r}_{ij}$ and $\vec{v}_{i}$ as the sum of all unit vectors from node $i$, $\vec{\omega}_{ji}$ denotes the vector rejection $\text{Rej}_{\vec{u}_{ij}}(\vec{v}_{i})$, which represents the vector component of $\vec{v}_{i}$ perpendicular to $\vec{u}_{ij}$.
 $\Theta_{i}$ denotes the sum of cosine values of $\alpha_{jik}$, which denotes the angle formed by node i and two of its neighboring nodes $i,j$. 
For the sum of dihedral angle's cosine values, RGC is expressed as:
\begin{align}
&\vec{v}_{i} = \sum_{j=1}^{N_{i}} \vec{u}_{ij},  \quad \vec{\omega}_{ij} = \text{Rej}_{\vec{u}_{ij}}(\vec{v}_{i}) = \vec{v}_{i} - \left\langle \vec{v}_{i}, \vec{u}_{ij} \right\rangle \cdot \vec{u}_{ij},\\
&\Phi_{(i,j)}  = \sum_{j=1}^{N_{i}} \sum_{k=1}^{N_{i}} \cos \varphi _{mijn} = \left\langle \vec{\omega}_{ij}, \vec{\omega}_{ji} \right\rangle,
\end{align}
where, the direction unit $\vec{v}_{i}$ is the sum of all unit vectors from node $i$ to its all neighboring nodes $j$, $\vec{\omega}_{ji}$ denotes the vector rejection $\text{Rej}_{\vec{u}_{ij}}(\vec{v}_{i})$, which represents the vector component of $\vec{v}_{i}$ perpendicular to $\vec{u}_{ij}$, termed as the vector rejection. 
Take $\Theta$ and $\Phi$ as target ,we design two loss functions based on the sum of cosine values of bond angle and dihedral angle as shown in Section~\ref{tab:ourmethod}.
%The specific calculation details of $f_{\text{angle}}$ and $f_{\text{dihedral}}$ are as follows:
%\begin{align}
%f_{\text{angle}}&=\text{MLP}(\text{Concat}(h_{i}^{L}, \text{Mean}(h_{j}^{L}, h_{k}^{L}))), \\
%f_{\text{dihedral}}&=\text{MLP}(\text{Concat}(\text{Mean}(h_{i}^{L}, h_{j}^{L}), \text{Mean}(h_{k}^{L}, h_{m}^{L}))).
%\end{align}

\label{appendix:acceleration}

%\subsection{Meta Gradient Descent}
%\label{appendix:meta}

\subsection{Supplementary ablation study}

We further design an ablation study on PCQM4Mv2 to verify the impact of each module in 3D PGT. From the results in Table \ref{tab:ablation pre-training}, we can summarize that: a) The backbone which is combined with local message passing and global attention module has more obvious performance advantages. b) The single generative pre-training task only focuses on the reconstruction of a local descriptor. Combining them can bring more significant pre-training benefits. c). The effectiveness of searched adaptive weights has been clearly verified. 
%\vspace{-5pt}

\begin{table}[h]
	%	\small
	\caption{Ablation study on PCQM4Mv2. In the ``Backbone'' group, we compare the performance of different base architectures. In the ``Pre-task'' group, we compare the performance of the corresponding metric as the only generative task. In the ``Fusion'' group, we compare the performance of average fusion and automated fusion of the designed three pretext tasks in this paper.}
	\label{tab:ablation pre-training}
	\vspace{-5pt}
	\begin{tabular}{c|lc}
		\toprule
		&  Variant & PCQM4Mv2 Valid MAE \\
		\midrule
		\multirow{5}{*}{\textbf{Backbone}} & GNN & 0.0812\\
		& Transformer & 0.0807 \\
		& GraphTrans & 0.0789 \\
		& GPS  & 0.0764 \\
		& Searched architecture & 0.0762 \\
		\midrule
		\multirow{4}{*}{\textbf{Pre-task}} & Bond Length & 0.0807 \\
		& Bond Angle & 0.0811 \\
		& Dihedral Angle & 0.0813 \\
		& Total Energy & 0.0798 \\
		\midrule
		\multirow{2}{*}{\textbf{Fusion}} & Average& 0.0781 \\ 
		& Automated& \textbf{0.0762} \\
		\bottomrule
	\end{tabular}
\end{table}

\section{Experiments Details}
We report the detailed hyperparameters setup for pre-training in Table \ref{tab:pre-training_hyper}. At the same time, the hyperparameter search space during finetune in all downstream tasks is shown in Table \ref{tab:finetune_searchspace}. In addition, in order to facilitate the implementation of various GNN variants, we use the popular GNN library: PYG (Pytorch Geometric) (version 2.0.1). 
For pre-training on different datasets, we set the max epoch number as Table ~\ref{tab:pre-training_hyper} and stop pre-training when the validation loss does not improve for 10 consecutive epochs.

%\subsection{Parameter Details}
%\subsection{Implementation}

\begin{table*}[h]
\small
  \caption{Details of used datasets. The datasets on the upper Section~are used for pre-training, so they all contain 3D information, and some datasets contain multiple conformers. The datasets in the middle Section~are 8 molecular property prediction tasks (classification) in the mainstream research line, the input of these downstream tasks is 2D molecular graph and the metric is AUC.}
  \label{tab:datasets}
  \begin{tabular}{lcccccc}
    \toprule
    Dataset & \#Molecules & \#Avg. atoms & \#Avg. bonds & 3D Information & Multi-conf. & Task \\
    \midrule
    PCQM4Mv2 & 3,746,619 & 14.1 & 14.6 & \Checkmark & \XSolidBrush & Regression \\
    QM9 & 130,831 & 18.0 & 18.6 & \Checkmark & \XSolidBrush & Regression \\
    GEOM-DRUGS & 304,293 & 44.4 & 46.4 & \Checkmark & \Checkmark & Regression \\
    \midrule
    BBBP & 2,039 & 24.1 & 26.0 & \XSolidBrush & \XSolidBrush  & Classification  \\
    Tox21 & 7,831 & 18.6 & 19.3 & \XSolidBrush & \XSolidBrush & Classification  \\
    ToxCast & 8,576 & 18.8 & 19.3 & \XSolidBrush & \XSolidBrush & Classification  \\
    Sider & 1,427 & 33.6 & 35.4 & \XSolidBrush & \XSolidBrush & Classification  \\
    MUV & 93,087 & 26.4 & 28.3 & \XSolidBrush & \XSolidBrush & Classification  \\ 
    HIV & 41,127 & 25.5 & 27.5 & \XSolidBrush & \XSolidBrush  & Classification \\ 
    Bace & 1,512 & 34.1 & 36.9 & \XSolidBrush & \XSolidBrush & Classification \\
    ClinTox & 1,477 & 26.2 & 27.9 & \XSolidBrush & \XSolidBrush & Classification  \\
    \midrule
    ESOL & 1,128 & 13.3 & 13.7 & \XSolidBrush & \XSolidBrush & Regression  \\
    LIPO & 4,200 & 27.0 & 29.5 & \XSolidBrush & \XSolidBrush & Regression  \\
    Malaria & 9,999 & 29.4 & 32.7 & \XSolidBrush & \XSolidBrush & Regression  \\
    CEP & 29,978 & 25.8 & 27.2 & \XSolidBrush & \XSolidBrush & Regression  \\
    \bottomrule
  \end{tabular}
\end{table*}

\begin{table*}
\caption{Hyperparameter setup for pre-training.}
\label{tab:pre-training_hyper}
\begin{tabular}{l|c|c|c}
\toprule
Hyperparameter & QM9 & GEOM-Drugs & PCQM4Mv2 \\
\midrule
Num of Layers & 5 & 5 & 10 \\
Hidden Dimension & 300 & 300 & 256 \\
Dropout & 0.1 & 0.1 & 0.2 \\
Attention Dropout & 0.05 & 0.05 & 0.1 \\
Batch Size & 128 & 128 & 256 \\
Num of heads & 16 & 16 & 32 \\
Readout & MEAN & MEAN & MEAN \\
Readout MLP Layers & 2 & 2 & 2 \\
Weight decay & 1e-5 & 1e-5 & 1e-5 \\
Peak learning rate & 1e-4 & 1e-4 & 1e-4 \\
Epochs & 100 & 100 & 150 \\
Warmup Epochs & 5 & 5 & 10 \\
Activation function & GELU & GELU & GELU \\ 
Learning rate decay & Cosine & Cosine & Cosine \\
\bottomrule
\end{tabular}
\end{table*}

\begin{table*}
\small
\caption{Hyperparameter search space during finetune.}
\label{tab:finetune_searchspace}
\begin{tabular}{lc}
\toprule
Hyperparameter & Search Space \\
\midrule
Learning rate & [5e-5, 1e-5, 5e-4, 1e-4, 5e-3, 1e-3, 5e-2] \\ 
Batch Size & [32, 64, 128, 256, 512] \\
Epochs & [60, 80, 100, 120, 150] \\ 
Dropout Ratio & [0.0, 0.1, 0.2, 0.3] \\
Attention Dropout & [0.0, 0.03, 0.05, 0.01] \\ 
\bottomrule
\end{tabular}
\end{table*}

\end{document}